\documentclass[showpacs,aps,graphicx,twocolumn]{revtex4}
\usepackage{graphicx}

\begin{document}

\title{Efficient electronic entanglement concentration assisted with single mobile electron}
\author{Yu-Bo Sheng,$^{1,2,4}$\footnote{Email address:
shengyb@njupt.edu.cn} Lan Zhou,$^3$, Wei-Wen Cheng,$^{1,2,4}$
Long-Yan Gong,$^{1,2,4}$ and Sheng-Mei Zhao,$^{1,4}$ }
\address{$^1$ Institute of Signal Processing  Transmission, Nanjing
University of Posts and Telecommunications, Nanjing, 210003,  China\\
$^2$College of Telecommunications \& Information Engineering,
Nanjing University of Posts and Telecommunications,  Nanjing,
210003,
China\\
 $^3$Beijing National Laboratory for Condensed Matter Physics, Institute of Physics,\\
Chinese Academy of Sciences, Beijing 100190, China\\
 $^4$Key Lab of Broadband Wireless Communication and Sensor Network
 Technology,
 Nanjing University of Posts and Telecommunications, Ministry of
 Education, Nanjing, 210003,
 China\\}
\date{\today}

\begin{abstract}
We present an efficient entanglement concentration protocol (ECP)
for mobile electrons with charge detection. This protocol is quite
different from other ECPs for one can obtain a maximally entangled
pair from a pair of less-entangled state and a single mobile
electron with a certain probability.  With the help of charge
detection, it can be repeated to reach a higher success probability.
It also does not need to know the coefficient of the original
less-entangled states. All these advantages may make this protocol
 useful in current distributed quantum information processing.
\end{abstract}
 \maketitle Entanglement plays
an important role in the current quantum communication
~\cite{rmp,Ekert91,teleportation,cteleportation,QSS1,QSS2,QSS3,QSDC1,QSDC2,QSDC3,densecoding,QSTS1,QSTS2,QSTS3}
and distributed quantum information processing field
~\cite{computation1,distributed1,distributed2,distributed3,distributed4}.
For most of the practical quantum communication and computation
protocols, people need to share a maximally entangled state with
each other. However, the entanglement resource is fragile, for the
maximally entangled state may be degraded into a mixed state or
become a less-entangled state when it interacts with the noisy
environment. People usually resort to the entanglement purification
~\cite{C.H.Bennett1,D.Deutsch,Pan1,shengpra,Pan2,lixhepp,dengonestep1,dengonestep2,wangchuan1,wangchuan2,wangchuan3,vanloock,feng,shengonestep}
to increase the fidelity of the mixed state and the entanglement
concentration~\cite{C.H.Bennett2,swapping1,Yamamoto1,zhao1,shengpra2,shengqic,shengpla,kim,wangxb,zhao2,Yamamoto2,dengnonsuo,shengpra3}
which will be detailed here to recover the less-entangled state to a
maximally entangled state. Currently, most of the protocols for
purification and concentration are focused on optical systems
~\cite{Pan1,Pan2,shengpra,Yamamoto1,zhao1,shengpra2,shengqic,kim,wangxb,zhao2,Yamamoto2,lixhepp,dengonestep1,shengonestep,
dengonestep2,wangchuan2,wangchuan3,vanloock}, for during the
transmission, the photons have weak interaction with the
environment.

On the other hand, there is another candidate for the flying qubit,
that is the  mobile electron. A strong interaction between different
electrons makes them feasible to interact flying electron spins with
other solid electron spins, since the coulomb interaction between
each electrons is strongly screened. In the recent years, the
investigation about the flying electron qubits becomes an active
study area
~\cite{Beenakker,costa,ciccarello1,ciccarello2,habgood,mastuzaki,Trauzettel}.
In 2004, Beenakker \emph{et al.} broke through the obstacle of the
no-go theorem with the help of charge detection and constructed a
controlled-NOT(CNOT) gate using beam splitters and spin rotations
near deterministically ~\cite{Beenakker}.
 With the help of charge detector,
people can construct the charge qubit ~\cite{Trauzettel}, perform
entanglement purification ~\cite{feng}, construct entangled spins
~\cite{Ionicioiu}, and prepare a multipartite entanglement analyzer
and cluster states ~\cite{Zhang}. Especially, the flying qubit can
also be used in the one dimensional system and create entanglement
between two distant matter qubits. Recently, Matsuzaki and Jefferson
proposed a protocol for distributed quantum information processing
with mobile electrons~\cite{mastuzaki}. In their protocol, they used
mobile electron spins as the mediators of the interaction between
the static qubits at each node and finally created a high quality
entanglement between each node. Unfortunately, the distributed
quantum entanglement may also be degraded into the less-entangled
state when it is coupled with the noisy environment. On the other
hand, with current technology, it is difficult to operate each
flying quibts perfectly, which can also lead the ideal maximally
entangled state to be degraded~\cite{mastuzaki}.

\begin{figure}
\includegraphics[width=8cm,angle=0]
{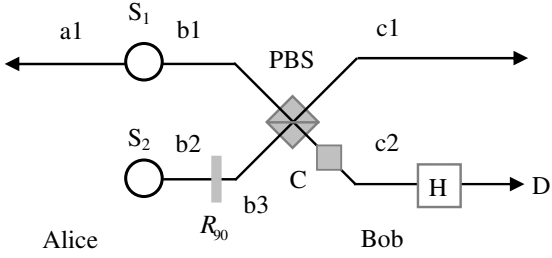} \caption{The schematic drawing of the principle of our ECP
based on charge detection. Only one pair of less-entangled state and
a single electron are required here. PBS can fully transmit spin up
$|\uparrow\rangle$ and reflect spin down $|\downarrow\rangle$. The
charge detector (C) can distinguish the charge number 1 from 0 and
2, but can not distinguish the number 0 and 2. D is the detector.}
\label{fig.1}
\end{figure}

Entanglement concentration is a powerful tool to recover a
less-entangled state into a maximally entangled one with only local
operation and classical communication.
 The first entanglement concentration protocol (ECP) was proposed by
Bennett \emph{et al.} in 1996 ~\cite{C.H.Bennett2}. This method is
called Schmidt projection method, in which they need the collective
measurement and need to know the exact coefficient of the initial
entangled state. In 2001, Zhao \emph{et al.} and Yamamoto \emph{et
al.} proposed two similar simplified ECPs with linear optical
elements based on the Schmidt projection method respectively. Both
ECPs have been realized experimentally ~\cite{zhao2,Yamamoto2}. In
2009, we proposed an ECP based on electrons ~\cite{shengpla}. Almost
all the current ECPs need two pairs of less-entangled states, but
after performing each ECP, at most one pair of maximally entangled
state can be obtained with a certain success probability.

Actually, using two copies of less-entangled pairs to obtain one
maximally entangled pair is not the optimal way. In this paper, we
show that we can perform the ECP with the same success probability
by using only one pair of less-entangled state and a single
electron. Compared with the previous ECPs, this protocol requires
less less-entangled resources. Moreover, analogized with the Ref.
~\cite{shengpla}, we adopt the charge detectors and polarization
beam splitter (PBS) to reconstruct our protocol, which makes it have
a higher success probability than those protocols with linear
optics. This protocol can also be used to concentrate the
multipartite entangled system.  All these advantages may make this
protocol more useful in current quantum communication and
distributed quantum information processing.

Before we start to explain our ECP, we first introduce the charge
detector (C) which is a key element shown in Fig. 1. The charge
detector can distinguish the occupation number 1 from the occupation
number 0 and 2. However, it can not distinguish between the
occupation number 0 and 2 ~\cite{Beenakker}, so in both two cases,
we define the charge detector will show 0 for simple. In Fig.1, we
suppose the source $S_{1}$ emits a pair of less-entangled state of
the form

\begin{eqnarray}
|\Phi\rangle_{a1b1}=\alpha|\uparrow\rangle_{a1}|\uparrow\rangle_{b1}+\beta|\downarrow\rangle_{a1}|\downarrow\rangle_{b1}.
\end{eqnarray}
Here $|\alpha|^{2}+|\beta|^{2}=1$.  $|\uparrow\rangle$ and
$|\downarrow\rangle$ are spin up and spin down respectively.
Meanwhile, the source $S_{2}$   emits a single electron to Bob of
the form
\begin{eqnarray}
|\Phi\rangle_{b2}=\alpha|\uparrow\rangle_{b2}+\beta|\downarrow\rangle_{b2}.\label{single}
\end{eqnarray}
Bob receives two electrons from the spatial modes  $b1$ and $b2$ and
Alice only receives one electron from  $a1$. Bob first performs a
bit-flip operation and makes $|\Phi\rangle_{b2}$ become
\begin{eqnarray}
|\Phi\rangle_{b2}\rightarrow|\Phi\rangle_{b3}=\alpha|\downarrow\rangle_{b3}+\beta|\uparrow\rangle_{b3}.
\end{eqnarray}

Therefore, the original entangled state of the three-electron state
can be written as
\begin{eqnarray}
|\Psi\rangle&=&|\Phi\rangle_{a1b1}\otimes|\Phi\rangle_{b3}\nonumber\\
&=&(\alpha|\uparrow\rangle_{a1}|\uparrow\rangle_{b1}+\beta|\downarrow\rangle_{a1}|\downarrow\rangle_{b1})
\otimes(\alpha|\downarrow\rangle_{b3}+\beta|\uparrow\rangle_{b3})\nonumber\\
&=&\alpha^{2}|\uparrow\rangle_{a1}|\uparrow\rangle_{b1}|\downarrow\rangle_{b3}+\beta^{2}|\downarrow\rangle_{a1}|\downarrow\rangle_{b1}|\uparrow\rangle_{b3}\nonumber\\
&+&\alpha\beta(|\uparrow\rangle_{a1}|\uparrow\rangle_{b1}|\uparrow\rangle_{b3}+|\downarrow\rangle_{a1}|\downarrow\rangle_{b1}|\downarrow\rangle)_{b3}.
\end{eqnarray}

Then Bob lets his two electrons pass through the PBS, which fully
transmits $|\uparrow\rangle$  and reflects $|\downarrow\rangle$. The
$|\Psi\rangle$ becomes
\begin{eqnarray}
|\Psi\rangle&\rightarrow&|\Psi\rangle'=\alpha^{2}|\uparrow\rangle_{a1}|\uparrow\rangle_{c2}|\downarrow\rangle_{c2}+\beta^{2}|\downarrow\rangle_{a1}|\downarrow\rangle_{c1}|\uparrow\rangle_{c1}\nonumber\\
&+&\alpha\beta(|\uparrow\rangle_{a1}|\uparrow\rangle_{c1}|\uparrow\rangle_{c2}+|\downarrow\rangle_{a1}|\downarrow\rangle_{c1}|\downarrow\rangle_{c2}).
\end{eqnarray}

From above equation, one can find that the item
$|\uparrow\rangle_{a1}|\uparrow\rangle_{c2}|\downarrow\rangle_{c2}$
means that the two electrons in Bob's location are both in the
spatial mode $c2$ while the item
$|\downarrow\rangle_{a1}|\downarrow\rangle_{c1}|\uparrow\rangle_{c1}$
means that the two electrons are both in the mode $c1$. However,
both the items
$|\uparrow\rangle_{a1}|\uparrow\rangle_{c1}|\uparrow\rangle_{c2}$
and
$|\downarrow\rangle_{a1}|\downarrow\rangle_{c1}|\downarrow\rangle_{c2}$
mean that the two electrons are in the modes $c1$ and $c2$,
respectively. Therefore, the charge detector will show 0 if the
original state collapses to
$|\uparrow\rangle_{a1}|\uparrow\rangle_{c2}|\downarrow\rangle_{c2}$
or
$|\downarrow\rangle_{a1}|\downarrow\rangle_{c1}|\uparrow\rangle_{c1}$,
And it will show 1 if the original state collapses to
\begin{eqnarray}
|\Psi\rangle''=\frac{1}{\sqrt{2}}(|\uparrow\rangle_{a1}|\uparrow\rangle_{c1}|\uparrow\rangle_{c2}
+|\downarrow\rangle_{a1}|\downarrow\rangle_{c1}|\downarrow\rangle_{c2}).\label{threephoton}
\end{eqnarray}

The probability of obtaining the state of Eq. (\ref{threephoton}) is
$2|\alpha\beta|^{2}$. Obviously, it is the three-electron maximally
entangled state. It is easy to get a two-electron maximally
entangled state from Eq. (\ref{threephoton}). Bob needs to perform a
Hadamard operation on his electron in the mode $c2$. It makes
\begin{eqnarray}
|\uparrow\rangle_{c2}\rightarrow\frac{1}{\sqrt{2}}(|\uparrow\rangle_{c2}+|\downarrow\rangle_{c2}),\nonumber\\
|\downarrow\rangle_{c2}\rightarrow\frac{1}{\sqrt{2}}(|\uparrow\rangle_{c2}-|\downarrow\rangle_{c2}).
\end{eqnarray}

After Bob performing the Hadamard operation, Eq.
(~\ref{threephoton}) becomes
\begin{eqnarray}
|\Psi\rangle''&\rightarrow&\frac{1}{2}[|\uparrow\rangle_{a1}|\uparrow\rangle_{c1}(|\uparrow\rangle_{c2}+|\downarrow\rangle_{c2})\nonumber\\
&+&|\downarrow\rangle_{a1}|\downarrow\rangle_{c1}(|\uparrow\rangle_{c2}-|\downarrow\rangle_{c2})]\nonumber\\
&=&\frac{1}{2}[(|\uparrow\rangle_{a1}|\uparrow\rangle_{c1}+|\downarrow\rangle_{a1}|\downarrow\rangle_{c1})|\uparrow\rangle_{c2}\nonumber\\
&+&(|\uparrow\rangle_{a1}|\uparrow\rangle_{c1}-|\downarrow\rangle_{a1}|\downarrow\rangle_{c1})|\downarrow\rangle_{c2}].
\end{eqnarray}

Then the last step for Bob is to measure the spin in the basis
$Z=\{|\uparrow\rangle,|\downarrow\rangle \}$. From above equation,
if the measurement result is $|\uparrow\rangle_{c2}$, the electron
pair in the modes $a1c1$  will become
\begin{eqnarray}
|\phi^{+}\rangle_{a1c1}=\frac{1}{\sqrt{2}}(|\uparrow\rangle_{a1}|\uparrow\rangle_{c1}+|\downarrow\rangle_{a1}|\downarrow\rangle_{c1}).
\end{eqnarray}

If the measurement result is $|\downarrow\rangle_{c2}$, the electron
pair in the modes $a1c1$ will become
\begin{eqnarray}
|\phi^{-}\rangle_{a1c1}=\frac{1}{\sqrt{2}}(|\uparrow\rangle_{a1}|\uparrow\rangle_{c1}-|\downarrow\rangle_{a1}|\downarrow\rangle_{c1}).
\end{eqnarray}
 Bob only needs to perform a phase-flip operation on his electrons to
get the $|\phi^{+}\rangle_{a1c1}$. If they share the pair
$|\phi^{\pm}\rangle_{a1c1}$, Bob will tell Alice that the protocol
is successful and asks Alice to retain her electron. In this way,
they can share a maximally entangled state from a less-entangled
state with the success probability of $2|\alpha\beta|^{2}$.

From above description, Bob chooses the case that the charge
detector's result is 1 and discards the case of 0. It is essentially
the partially parity check gate which picks up the even parity
states $|\uparrow\rangle|\uparrow\rangle$ and
$|\downarrow\rangle|\downarrow\rangle$ but discards the odd parity
states $|\uparrow\rangle|\downarrow\rangle$ and
$\downarrow\rangle|\uparrow\rangle$. In this case, the function of
PBS is similar as it is in the optical systems
~\cite{zhao1,Yamamoto1}. In Refs. ~\cite{zhao1,Yamamoto1}, they need
to check that both the output modes of the optical PBS  contain
exactly and only contain one photon. It is so called the
post-selection principle. Therefore, even if they successfully
perform their ECPs, the maximally entangled state would be destroyed
by the single photon detector. The maximally entangled photon pair
can not be remained for further application. In this protocol, Bob
can judge the successful case from the charge detection results. The
charge qubit carries both the spin degree of freedom and the charge
degree of freedom. As charge and spin are commute and a measurement
of charge leaves the spin qubit unaffected, the charge detection
does not affect the entangled state of the electrons.

Actually, the charge detector and PBS are more powerful than it has
been described above, for the discarded items  are essentially the
lesser-entangled state which
 can be reconcentrated in a second step. That is to say, we can not only pick up the even parity states, but also pick up the odd
 parity states.
In Fig. 2, we add another PBS say PBS$_{2}$ to reconstruct our ECP.
We denote the whole setup P gate shown in Fig. 2. If the charge
detector's result is 0, the original state will collapse to
\begin{eqnarray}
|\Psi\rangle'''=\alpha^{2}|\uparrow\rangle_{a1}|\uparrow\rangle_{c2'}|\downarrow\rangle_{c2'}+\beta^{2}|\downarrow\rangle_{a1}|\downarrow\rangle_{c1'}|\uparrow\rangle_{c1'}
\end{eqnarray}

\begin{figure}
\includegraphics[width=9cm,angle=0]
{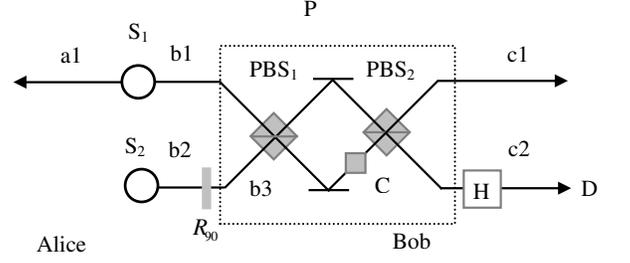} \caption{The schematic drawing of the principle of
reconstructing of our ECP. If the measurement result of charge
detector is 0, the remaining lesser-entangled pair can also be
reused to perform the entanglement concentration. Another PBS says
PBS$_2$ is used to couple the state into the same spatial mode. P
denotes that it plays essentially the role of the parity check
gate.} \label{fig.2}
\end{figure}

It means that after passing through the PBS$_{1}$ , both the two
electrons are in the same spatial mode, while with the help of
PBS$_{2}$, they are coupled into

\begin{eqnarray}
|\Psi\rangle'''=\alpha^{2}|\uparrow\rangle_{a1}|\uparrow\rangle_{c1}|\downarrow\rangle_{c2}+\beta^{2}|\downarrow\rangle_{a1}|\downarrow\rangle_{c1}|\uparrow\rangle_{c2}.
\end{eqnarray}
It means that the two electrons in Bob's location are in the
different spatial modes, say $c1$ and $c2$, respectively. Then after
performing the Hadamard operation and measuring the spin of the
electron on the mode $c2$ , they can get
\begin{eqnarray}
|\Phi\rangle_{1}=\alpha^{2}|\uparrow\rangle_{a1}|\uparrow\rangle_{c1}+\beta^{2}|\downarrow\rangle_{a1}|\downarrow\rangle_{c1},\label{less1}
\end{eqnarray}
if the measurement result is $|\uparrow\rangle$. They  can get
\begin{eqnarray}
|\Phi\rangle_{2}=\alpha^{2}|\uparrow\rangle_{a1}|\uparrow\rangle_{c1}-\beta^{2}|\downarrow\rangle_{a1}|\downarrow\rangle_{c1},\label{less2}
\end{eqnarray}
if the measurement result is $|\downarrow\rangle$. Both Eqs.
(\ref{less1}) and (\ref{less2}) are lesser-entangled states. They
can be reconcentrated in the next step. Briefly speaking, if they
obtain $|\Phi\rangle_{1}$, Bob needs to choose another single
electron of the form
\begin{eqnarray}
|\Phi\rangle'_{b2}=\alpha^{2}|\uparrow\rangle_{b2}+\beta^{2}|\downarrow\rangle_{b2}.
\end{eqnarray}
After rotating it with $R_{90}$, it becomes
\begin{eqnarray}
|\Phi\rangle'_{b3}=\alpha^{2}|\downarrow\rangle_{b3}+\beta^{2}|\uparrow\rangle_{b3}.
\end{eqnarray}
Then the three-electron system evolves as
\begin{eqnarray}
|\Phi\rangle_{2}&\otimes&|\Phi\rangle'_{b3}=(\alpha^{2}|\uparrow\rangle_{a1}|\uparrow\rangle_{c1}+\beta^{2}|\downarrow\rangle_{a1}|\downarrow\rangle_{c1})\nonumber\\
&\otimes&(\alpha^{2}|\downarrow\rangle_{b3}+\beta^{2}|\uparrow\rangle_{b3})\nonumber\\
&=&\alpha^{4}|\uparrow\rangle_{a1}|\uparrow\rangle_{c1}|\downarrow\rangle_{b3}+\beta^{4}|\downarrow\rangle_{a1}|\downarrow\rangle_{c1}|\uparrow\rangle_{b3}\nonumber\\
&+&\alpha^{2}\beta^{2}(|\uparrow\rangle_{a1}|\uparrow\rangle_{c1}|\uparrow\rangle_{b3}+|\downarrow\rangle_{a1}|\downarrow\rangle_{c1}|\downarrow\rangle_{b3}).\nonumber\\
\end{eqnarray}

Obviously, if the charge detection C=1, they can obtain the same
three-electron state as described in Eq. (~\ref{threephoton}) with a
probability of $2|\alpha\beta|^{4}$. Otherwise, if C=0, the remained
state is a  lesser-entangled state and  can be reconcentrated in a
third round. In this way, they can repeat this protocol to get a
higher success probability than other protocols.

\begin{figure}
\includegraphics[width=8cm,angle=0]
{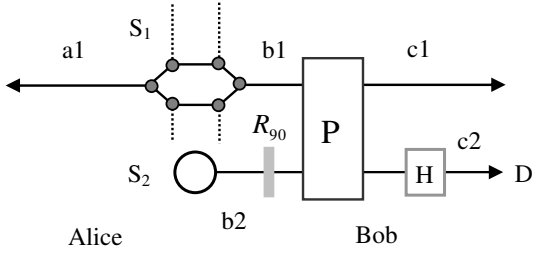} \caption{Schematic diagram of the multipartite entanglement
concentration protocol. $N$ particles in the multipartite GHZ state
from the source $S_{1}$ are sent to to $N$ parties, say, Alice, Bob,
Charlie, etc. The source $S_{2}$ also emits a single electron to
Bob. The P is the parity check gate shown in Fig.2. It comprises the
PBS$_{1}$, charge detector C and PBS$_{2}$. Only Bob needs to
perform this concentration.} \label{fig.3}
\end{figure}

It is straightforward to extend  this protocol to multi-partite pure
entangled state systems. An $N$-electron less-entangled system can
be described as
\begin{eqnarray}
|\Phi\rangle_{N}=\alpha|\uparrow\rangle_{1}|\uparrow\rangle_{2}\cdots|\uparrow\rangle_{N}+\beta|\downarrow\rangle_{1}|\downarrow\rangle_{2}\cdots|\downarrow\rangle_{N}.\label{Nparticle}
\end{eqnarray}
The $N$ electrons are emitted from $S_{1}$ and sent to $N$ parties,
say, Alice, Bob, Charlie, etc, as shown in Fig. 3.  Alice gets the
electron of number 1 in the spatial mode $a1$. Bob gets number 2 in
the spatial mode $b1$ and Charlie gets the number 3, etc. The source
of $S_{2}$ also emits a single electron to Bob with the same form of
Eq. (~\ref{single}). After rotating it by $90^{\circ}$, the
composite system can be described as

\begin{eqnarray}
&&|\Psi\rangle_{N+1}=|\Phi\rangle_{N}\otimes|\Phi\rangle_{b3}=(\alpha|\uparrow\rangle_{1}|\uparrow\rangle_{2}\cdots|\uparrow\rangle_{N}\nonumber\\
&&+\beta|\downarrow\rangle_{1}|\downarrow\rangle_{2}\cdots|\downarrow\rangle_{N})\otimes(\alpha|\downarrow\rangle_{b3}+\beta|\uparrow\rangle_{b3})\nonumber\\\label{N+1particle}
&&=\alpha^{2}|\uparrow\rangle_{1}|\uparrow\rangle_{2}|\downarrow\rangle_{b3}\cdots|\uparrow\rangle_{N}\nonumber\\
&&+\beta^{2}|\downarrow\rangle_{1}|\downarrow\rangle_{2}|\uparrow\rangle_{b3}\cdots|\downarrow\rangle_{N})\nonumber\\
&&+\alpha\beta(|\uparrow\rangle_{1}|\uparrow\rangle_{2}|\uparrow\rangle_{b3}\cdots|\uparrow\rangle_{N}
+|\downarrow\rangle_{1}|\downarrow\rangle_{2}|\downarrow\rangle_{b3}\cdots|\downarrow\rangle_{N}).\nonumber\\\label{multi}
\end{eqnarray}

Subsequently, the electrons in the spatial modes $b1$ and $b3$ in
Bob's location pass through the $P$ gate.  If the charge detector's
result is $C=1$, the Eq. (~\ref{multi}) will collapse to
\begin{eqnarray}
|\Psi'\rangle_{N+1}&=&\frac{1}{\sqrt{2}}(|\uparrow\rangle_{1}|\uparrow\rangle_{2}|\uparrow\rangle_{c2}\cdots|\uparrow\rangle_{N}\nonumber\\
&+&|\downarrow\rangle_{1}|\downarrow\rangle_{2}|\downarrow\rangle_{c2}\cdots|\downarrow\rangle_{N}),
\end{eqnarray}
with the probability of $2|\alpha\beta|^{2}$. Finally, Bob performs
a Hardamard operation, and measures the electron in the  mode $c2$
in the basis $Z=\{|\uparrow\rangle,|\downarrow\rangle\}$. If the
measurement result is $|\uparrow\rangle$, they will get
\begin{eqnarray}
|\Psi'\rangle_{N}&=&\frac{1}{\sqrt{2}}(|\uparrow\rangle_{1}|\uparrow\rangle_{2}\cdots|\uparrow\rangle_{N}\nonumber\\
&+&|\downarrow\rangle_{1}|\downarrow\rangle_{2}\cdots|\downarrow\rangle_{N}),\label{multi1}
\end{eqnarray}
and if the measurement result is $|\downarrow\rangle$, they will get
\begin{eqnarray}
|\Psi''\rangle_{N}&=&\frac{1}{\sqrt{2}}(|\uparrow\rangle_{1}|\uparrow\rangle_{2}\cdots|\uparrow\rangle_{N}\nonumber\\
&-&|\downarrow\rangle_{1}|\downarrow\rangle_{2}\cdots|\downarrow\rangle_{N}).\label{multi2}
\end{eqnarray}
Both Eqs. (~\ref{multi1}) and (~\ref{multi2}) are the $N$-electron
maximally entangled states.

Otherwise, if the charge detector's result is $C=0$, then Eq.
(\ref{multi}) becomes
\begin{eqnarray}
|\Psi''\rangle_{N+1}&=&\alpha^{2}|\uparrow\rangle_{1}|\uparrow\rangle_{2}|\downarrow\rangle_{b3}\cdots|\uparrow\rangle_{N}\nonumber\\
&+&\beta^{2}|\downarrow\rangle_{1}|\downarrow\rangle_{2}|\uparrow\rangle_{b3}\cdots|\downarrow\rangle_{N}.
\end{eqnarray}
After performing the Hadamard operation and measuring the electron
in c2 mode in the $Z$ basis, the above state becomes
\begin{eqnarray}
|\Psi^{\pm}\rangle_{N}&=&\alpha^{2}|\uparrow\rangle_{1}|\uparrow\rangle_{2}\cdots|\uparrow\rangle_{N}\nonumber\\
&\pm&\beta^{2}|\downarrow\rangle_{1}|\downarrow\rangle_{2}\cdots|\downarrow\rangle_{N}.
\end{eqnarray}
Compared with Eq. (~\ref{Nparticle}), it is also a multipartite
less-entangled state which can be reconcentrated into a maximally
entangled state. '+' or '-' is decided by the measurement result
$|\uparrow\rangle$ or $|\downarrow\rangle$, respectively.

So far, we have fully described our ECP. It is interesting to
compare this protocol with Ref. ~\cite{shengpla}. In Ref.
~\cite{shengpla}, we resort two copies of less-entangled pairs to
perform the concentration. We can get one pair of maximally
entangled state with the success probability of
$2|\alpha\beta|^{2}$. In this protocol,
 we use only one pair of less-entangled state and a single electron which can reach
the same success probability with Ref. ~\cite{shengpla}. Moreover,
during the whole protocol, only one-way classical communication is
required. That is Alice only needs
 to receive the information from the Bob's measurement and to judge
 whether the protocol is successful or fail. If the protocol is a failure,
 Alice needs to do nothing. If the protocol is successful, Bob will
 tell Alice the remaining state is $|\phi^{+}\rangle$ or
 $|\phi^{-}\rangle$ according to his measurement. In previous protocols
~\cite{zhao1,Yamamoto1,shengpra,shengpla}, all of the parties have
to participate the whole procedure, to measure their electrons and
check their results to each other to judge the remained state if the
protocol is successful. So this protocol is much more simple,
especially when it is used to concentration multi-partite system,
for only one of the parties needs to perform the protocol and then
report his results to others. Finally, let us briefly discuss the
key ingredient of this protocol here, that is charge detector
~\cite{buks,elz,shaner}. It has been realized in a two-dimensional
electron gas. It was reported that currently achievable time
resolution for charge detection is $\mu s$ ~\cite{elz}. In a
semiconductor it was reported that the resolution required for
ballistic electrons is less than 5 ps ~\cite{shaner}. Compared with
flying qubit, it may be more practical to use isolated electrons in
an array of quantum dots, as pointed by Ref. ~\cite{Beenakker}.

In conclusion, we have proposed an mobile electron ECP based on
charge detection. Compared with other protocols, it has several
advantages: first, it dose not require the post-selection principle,
and can be repeated to reach a higher efficiency than those based on
linear optical elements; second, only one pair of less-entangled
state and one-way classical communication are required, which make
this protocol more economic and simple than others.  All these
advantages make this ECP more useful in current quantum
communication and distributed quantum information processing.

\section*{ACKNOWLEDGEMENTS}
This work was supported by the National Natural Science Foundation
of China under Grant No. 11104159,  Scientific Research Foundation
of Nanjing University of Posts and Telecommunications under Grant
No. NY211008,  University Natural Science Research Foundation of
JiangSu Province under Grant No. 11KJA510002,  the open research
fund of Key Lab of Broadband Wireless Communication and Sensor
Network Technology (Nanjing University of Posts and
Telecommunications), Ministry of Education, China, and a Project
Funded by the Priority Academic Program Development of Jiangsu
Higher Education Institutions.

\end{document}